\documentclass[aps,prd,twocolumn,superscriptaddress]{revtex4-2}

\usepackage[utf8]{inputenc}

\usepackage{mathtools}
\usepackage{amsfonts}
\usepackage{mathrsfs}
\usepackage{bbm}
\usepackage{physics}
\usepackage{slashed}
\usepackage{tensor}

\usepackage{graphicx}
\usepackage{color, float}
\usepackage{array}
\usepackage[abs]{overpic}

\usepackage{placeins}

\usepackage{makecell}
\usepackage{subcaption}

\usepackage{xspace}
\usepackage{siunitx}
\usepackage{xfrac}
\usepackage{hyperref}
\usepackage[nameinlink]{cleveref}
\usepackage{appendix}

\usepackage{xifthen}
\usepackage{xcolor}
\hypersetup{
	colorlinks,
	linkcolor={red!75!black},
	citecolor={blue!75!black},
	urlcolor={blue!75!black}
}

\usepackage{booktabs}
\usepackage{multirow}

\newcolumntype{C}{>{$}c<{$}}
\AtBeginDocument{
	\heavyrulewidth=.08em
	\lightrulewidth=.05em
	\cmidrulewidth=.03em
	\belowrulesep=.65ex
	\belowbottomsep=0pt
	\aboverulesep=.4ex
	\abovetopsep=0pt
	\cmidrulesep=\doublerulesep
	\cmidrulekern=.5em
	\defaultaddspace=.5em
}

\newcommand{\gettitle}{Nonperturbative treatment of a quenched Langevin field theory}

\newcommand{\getHeidelbergAffiliation}{\affiliation{Institut f{\"u}r Theoretische Physik, Universit{\"a}t Heidelberg, Philosophenweg 16, 69120 Heidelberg, Germany}}
\newcommand{\getZuerichAffiliation}{\affiliation{Institut f{\"u}r Theoretische Physik, ETH Z{\"u}rich, Wolfgang-Pauli-Str. 27, 8093 Z{\"u}rich, Switzerland}}
\newcommand{\getKolnAffiliation}{\affiliation{Institut f\"ur Theoretische Physik, Universit\"at zu K\"oln, D-50937 Cologne, Germany}}
\newcommand{\getMainzAffiliation}{\affiliation{Institute of Physics, Johannes Gutenberg University Mainz, 55099 Mainz, Germany}}
\hypersetup{
	pdftitle={\gettitle},
	pdfauthor={Ihssen},
	pdfkeywords=
	{generalised RG flows} {renomalisation group} ,
	bookmarksopen=true,
	bookmarksopenlevel=2,
	bookmarksnumbered=true
}

\begin{document}	
	\title{\gettitle}
	
	\author{Friederike Ihssen}\getHeidelbergAffiliation\getZuerichAffiliation
	\author{Valerio Pagni}\getZuerichAffiliation
    \author{Jamir Marino}\getMainzAffiliation
    \author{Sebastian Diehl}\getKolnAffiliation
	\author{Nicolò Defenu}\getZuerichAffiliation
	
	\begin{abstract}
        We present a novel approach within the functional renormalization group framework for computing critical exponents that characterize the time evolution of out-of-equilibrium many-body systems.
        Our approach permits access to quantities involved in the renormalization procedure, using an expansion about time-translation invariant problems. This expansion can be upgraded to a fully time-dependent computation by iteration. As a prototypical example, we compute the aging exponent $\theta$ describing the dynamics of model A following a sudden quench to the critical point. Already at leading order, the approach demonstrates remarkable accuracy when compared with MC simulations and resummed perturbative expansions in the range $2<d<4$. This yields results that surpass those of the two-loop $\epsilon$ expansion in accuracy and match analytically known benchmarks at large $N$. These findings contribute to a deeper understanding of out-of-equilibrium universality and open new avenues for non-perturbative studies of critical dynamics, as well as for exploring the critical behavior of systems with spatial boundaries. 
	\end{abstract}

	\maketitle
 
\section{Introduction}

The emergence of universality in out-of-equilibrium systems is a major open question in many-body physics~\cite{tauber2017phase}. A key signature of universality is the appearance of self-similar scaling behavior, where dynamical observables evolve according to universal exponents. In quantum many-body systems, such behavior has been extensively investigated in diverse settings, including isolated systems undergoing unitary time evolution~\cite{polkovnikov2011colloquium}, long-range interacting models~\cite{monroe2021programmable,defenu2023long}, and dissipative open systems~\cite{mivehvar2021cavity,sieberer2023universality}. Thanks to the flexibility of modern quantum simulators and cold atom platforms, these studies have provided crucial insights into universality in non-equilibrium dynamics.

In general, non-equilibrium systems can be categorized into those undergoing time evolution from an initial state and those that reach non-equilibrium stationary states due to continuous driving and coupling to an external reservoir. In this work, we focus on the former: transient dynamics following a sudden change in a control parameter, where the system evolves in time and may eventually relax toward equilibrium.

The rapid growth of experimental studies has further motivated theoretical efforts to characterize universal properties of out-of-equilibrium dynamics, particularly the determination of critical exponents. However, despite significant progress, our understanding of universality in the non-equilibrium regime remains far less developed than in equilibrium statistical physics.

From a theoretical perspective, the calculation of dynamical scaling indices has been approached mostly through perturbative and strong-coupling expansions~\cite{aarts2002classical,chiocchetta2017dynamical,marino2022universality,preis2023stable}. In the scaling regime close to equilibration some non-perturbative calculations exist, see~\cite{canet2011general,canet2007non, batini2023dissipation} for applications to model A in particular. The determination of critical exponents far from equilibrium incites the use of advanced techniques, such as classical statistical simulations~\cite{berges2010dynamic, schweitzer2020spectral, schweitzer2022critical} or the functional renormalization group (fRG)~\cite{berges2009nonthermal, canet2011general, dupuis2021the, roth2023critical}. The non-perturbative nature of these quantities becomes apparent in low-dimensional systems, where most perturbative results break down.

Addressing genuine out-of-equilibrium universality in a non-perturbative manner is challenging due to the explicit breaking of time-translational invariance and the introduction of a 'temporal boundary.' This has historically hindered the application of functional truncations of the effective action to the out-of-equilibrium universality observed after sudden quenches at criticality. To overcome this issue, we perform an expansion of the time-dependent functional RG equations around their stationary (infinite time) solutions. In this approach, only the time-dependent part of the field is assumed to remain small throughout the dynamical evolution, while its stationary expectation is not constrained. 

To demonstrate our approach, we calculate the dynamical scaling indices of a finite-temperature field theory subjected to a sudden quench to its thermal critical point. These dynamics belong to the model A class within the traditional classification of out-of-equilibrium phase transitions~\cite{hohenberg1977theory} and can be used to study the critical dynamics of various physical systems, including ferromagnetic transitions, liquid-gas critical points, and potentially the chiral phase transition in QCD~\cite{tan2022real}. The order parameter $M$, representing the magnetization in an $O(N)$ system, evolves toward equilibrium after the initial temperature quench, exhibiting two distinct dynamical regimes: the short-time dynamics and the eventual approach to equilibrium.
The short-time critical dynamics is governed by the aging exponent $\theta$ or the initial slip exponent $\theta'$~\cite{janssen1989new,janssen1992renormalized}, which we introduce below and compute for general $O(N)$ theories. For a review of the aging exponent see~\cite{calabrese2005ageing}.

Our results converge up to $d=2$ without encountering the fundamental issues faced by perturbative approaches and successfully reproduce the analytically known large-$N$ benchmarks. The equilibrium input data is obtained within an LPA' truncation, utilizing state-of-the-art numerical techniques to solve the fixed-point equations. Specifically, we employ the Chebyshev spectral method, which provides the fully field-dependent equilibrium potential\,\cite{borchardt2015global}. Aging dynamics in $O(N)$ field theory align particularly well with the assumptions of our method, as the breaking of time-translational symmetry is confined to exponentially vanishing portions of the propagators\,\cite{chiocchetta2016universal}.

The paper is structured as follows: We begin by a short introduction to critical systems with a boundary and their connection to aging behavior in \Cref{sec:InitialSlip}. Next, we outline the functional RG, as well as the approximation used in \Cref{sec:Method}. \Cref{sec:HomogeneousExp} discusses the computation of time-dependent quantities in an iterative setup and represents the main advance of this work. Finally, we present results for the aging exponent in \Cref{sec:results} and close with a discussion in \Cref{sec:conclusion}.

\section{Critical systems with boundaries and aging dynamics}\label{sec:InitialSlip}

The present work considers the non-equilibrium dynamics ensuing from a sudden quench of parameters at the time surface specified by $t=t_0$. However, the techniques developed in \Cref{sec:HomogeneousExp} may also be used to study critical phenomena in systems with a \textit{spatial} boundary, i.e.~a spatial inhomogeneity. This concerns in particular the equilibrium behavior of semi-infinite systems with a spatial boundary.

We begin with a brief discussion to motivate the appearance of new critical exponents in systems with a boundary in \Cref{sec:Boundary}, where we also make the connection between systems with spatial and temporal surfaces. An introduction to model A follows in \Cref{sec:ModelA}.

\subsection{Boundary criticality and aging}\label{sec:Boundary}

Consider a system in $d$-dimensional space with a boundary, $[0,L]^d$, where $L$ is the size of the system. A many-body statistical system, e.g.~an $O(N)$ model, placed on $[0,L]^d$ exhibits surface effects whose influence penetrate the bulk according to the size of the correlation length $\xi$ \cite{diehl1997theory}. The finite value of $L$ gives rise to finite-size effects that can be neglected by taking the limit $L\to\infty$, in which case we are considering a semi-infinite system with a free boundary. In the thermodynamic limit we can have criticality in the bulk with $\xi\to\infty$. While the usual critical exponents describe the singular behavior of bulk observables such as (bulk) magnetization, it is found that there are novel exponents for the corresponding surface observables. From a field-theoretical perspective \cite{diehl1986field}, boundary critical exponents arise due to the additional ultraviolet singularities occurring when two field insertions are located at $r$ and $r'\approx r$, both on the boundary of the system.

Similarly, the presence of a boundary in time (rather than in space) can generate new independent critical exponents related to non-equilibrium dynamical scaling behavior \cite{janssen1989new,janssen1992renormalized}. 
The non-equilibrium short-time exponents are related to insertions of the field which are located on the temporal hyper-surface at $t=t_0$, in a clear analogy with the spatial counterpart of the problem. On longer time scales, the dynamics of relaxation to equilibrium \cite{hohenberg1977theory} is expected to dominate the scaling behavior. Finally, we notice that time-translation invariance is clearly broken because of the presence of the time boundary. 

Together with the property of non-exponential relaxation (or slow dynamics), the ingredients of dynamical scaling and breaking of time-translation invariance define ``aging" dynamics \cite{henkel2011non}. Because substantial cooperative effects and large fluctuations are typically responsible for aging, one necessarily needs to harness the power of non-mean field techniques such as the renormalization group to correctly capture such dynamical behavior.

While in this paper we are focusing on the dynamics following a critical quench as described in the next section, aging is often studied also in relation to the \textit{non}-critical dynamics of phase-ordering \cite{henkel2011non,bray1994theory}. Moreover, significant attention has been devoted to aging in spin glasses, see e.g.~\cite{henkel2007ageing,cugliandolo2002dynamics}.

\subsection{The initial slip exponent}\label{sec:ModelA}

We give here a brief reminder on the dynamics towards equilibrium of a non-conserved order parameter, also known as model A \cite{hohenberg1977theory}. Within this setting we provide a concrete application of the techniques developed later and determine the aging behavior of the system after a temperature quench. A more in-depth introduction to the subject can be found in \cite{calabrese2005ageing}.

Model A refers to the evolution of a $N$-component classical field $\varphi = (\varphi^1, \dots, \varphi^N)$ described by the Langevin equation
\begin{align}\label{eq:LangevinEQ}
	\dot  \varphi = -D \frac{\delta \mathcal{H}}{\delta \varphi} + \zeta \,,
\end{align}
where $D$ is a constant relaxation coefficient and $\zeta$ a zero-mean Markovian and Gaussian noise with correlation
\begin{align}
	\langle \zeta(r_1, t_1)\zeta(r_2, t_2) \rangle = 2 D \, \delta(x_1,x_2) \,,
\end{align}
and the space-time variable $x=(r,t)$. We are working in the units where $k_B T = 1$. The Hamiltonian $\mathcal{H}$ describes an $O(N)$ model and is given by
\begin{align}\label{eq:Hamiltonian}
	\mathcal{H} = \int_{r } \left\{ \frac12 (\nabla \varphi)^2 + \frac{\mu}{2} \varphi^2 + \frac{g}{8}(\varphi^2)^2 \right\} \,,
\end{align}
where $\int_{r} = \int d^d r$.

A large portion of the literature has been focusing only on the long-time relaxation of model A \cite{hohenberg1977theory,tauber2014critical}. Here, in contrast, we consider its out-of-equilibrium scaling at short times in analogy with \cite{janssen1989new}. We assume that the initial state is a high-temperature one, meaning that it has short-range correlations and a Gaussian theory is able to capture its non-critical physics. If we denote the inverse correlation length as $\mu_0$, then the value of $\mu_0$ is very large. The initial state is also prepared with a small magnetization $M_0>0$. The dynamics starts at $t=t_0$, when the temperature is suddenly quenched to its critical value $T=T_c$ and possible external fields are set to zero. At large times we observe relaxation to equilibrium.
Thus, to study this dynamics information about the initial state has to be taken into account.

Observables are computed by averaging over solutions of the stochastic differential equation \labelcref{eq:LangevinEQ}. One such example is the mean magnetization, whose evolution at the critical point can be described by \cite{janssen1989new}
\begin{align}
	M(t) = \langle \varphi(t)\rangle  = M_0 \, t^{\theta'} f_M\left(t^{\theta' +\beta /(\nu z)} M_0\right)\,,
\end{align}
with the scaling function
\begin{align}
	f_M (x)  \propto \begin{cases}
		1 \,, \quad & \text{if $x=0$} \\
		1/x \,, \quad & \text{if $x\to\infty$}
	\end{cases} \,.
\end{align}
The scaling exponent $\theta'$ is a universal scaling parameter known as the initial slip exponent. This is a non-equilibrium scaling parameter, which determines the behavior at (macroscopically) \textit{short} times after the initial quench, and is independent of the critical exponents at (or close to) equilibrium, which govern the late-time behavior of observables. In particular, the scaling forms of the magnetization, correlation function, correlation length, and relaxation time are captured by the exponents $\beta, \eta, \nu$, and $z$, respectively -- see e.g. \cite{cardy1996scaling,tauber2014critical} -- which have been studied in detail with the fRG in \cite{codello2013n, defenu2015truncation, codello2015critical,defenu2020fate}.

For completeness we also indicate the scaling of the dynamic susceptibility
\begin{align}\label{eq:susceptibility}
	\chi(t_1,t_2,q) = q^{z + \eta - 2} \left(\frac{t_1}{t_2}\right)^\theta f_\chi \left(q \xi, q^z t_1\right) \,,
\end{align}
valid as $t_2 \to t_0$, where $\xi$ denotes the correlation length and $f_\chi$ the corresponding scaling function. \Cref{eq:susceptibility} exemplifies aging in two-time correlation functions, where the dependence on $t_1$ and $t_2$ does not occur through the time-translation invariant combination $(t_1-t_2)$. 
The scaling of the dynamic susceptibility is given by the exponent $\theta$, which is related to $\theta'$ by
\begin{align}\label{eq:initial_slip_vs_aging}
	\theta' = \theta + (2 - z -\eta)/z  \,.
\end{align}
%

\section{Functional methods}\label{sec:Method}

The problem of solving the stochastic differential equation \labelcref{eq:LangevinEQ} can be reformulated in terms of a path integral using the MRSJD \cite{martin1973statistical, janssen1976on, de1978dynamics} or response field formalism. The central object of our computation is the effective action $\Gamma[\phi, \tilde \phi]$, which is given in terms of the expectation value $\phi = \langle \varphi \rangle$ of the classical field and the response field $\tilde \phi$.

For a more in-depth introduction to the MSRJD framework in the context of statistical mechanics in the fRG, we defer to \cite{canet2007non,canet2011general,dupuis2021the}. In the present section, we gather the relevant details for this work: \Cref{sec:Trunc} contains the notation and truncation of the effective action. \Cref{sec:fRG} briefly introduces the fRG within the response field formalism and gives some technical details of the implementation. 

\subsection{Truncation of the effective action}\label{sec:Trunc}
We focus on the $O(N)$ symmetric generalization of the model A, i.e.~kinetic $O(N)$ models, whose equilibrium universal properties have been subject of recent investigations \cite{duclut2017frequency}.
We consider the time evolution of a $N$-component field $\phi$ within the response field approach \cite{martin1973statistical, janssen1976on, de1978dynamics}, with
\begin{align}
	\phi = (\phi^1, \dots, \phi^N)^t \,, \quad
	\tilde \phi = (\tilde \phi^1, \dots, \tilde \phi^N)^t \,,
\end{align}
where $\tilde \phi$ is the response field.
In order to obtain the universal short time dynamics of the system we consider the effective action
\begin{align}
	\label{eq:ansatz1}
	&\Gamma_{k}[\phi,\tilde{\phi}]=\Gamma_{0}[\phi_0,\tilde{\phi}_0] \\[1ex]
	&+\int_t \theta(t-t_{0}) \int_{r}\tilde{\phi}\left(Z\partial_{t}\phi-K\nabla^2 \phi+U^{(1)}(\phi)-D\tilde{\phi}\right) \,,\nonumber
\end{align}
where the superscript $^{(n)}$ represents the $n$-th derivative with respect to $\phi$. The boundary action for a quench from the far high-temperature phase takes the form \cite{chiocchetta2016universal}
\begin{align}\label{eq:InitialC}
	&\Gamma_0[\phi_0, \tilde \phi_0] = \int_{r} W(\tilde \phi_0, \phi_0), \quad \text{where}
    \nonumber\\
    &W(\tilde \phi_0, \phi_0) = Z_0 \tilde{\phi}_0\phi_0 -\frac{Z_0^2}{2 \mu_0} \tilde{\phi}_0^2 \,,
\end{align}
where $\phi_{0}$ and $\tilde{\phi}_{0}$ are, respectively, the order parameter $\phi$ and the response field $\tilde{\phi}$ evaluated at the initial time $t_{0}$. As in \Cref{sec:Boundary}, $\mu_0$ describe short-range Gaussian correlations at the initial scale, while $Z_0$ is the renormalization of the boundary response field. The boundary action \labelcref{eq:InitialC} encodes the initial Gaussian distribution of the scalar field at $t=t_0$. In future investigations, we may generalize the form of 
$W(\tilde \phi_0, \phi_0)$ in \labelcref{eq:InitialC} by including higher orders terms in the boundary fields, or even treat it functionally, without truncations.

\subsection{Functional RG equations}\label{sec:fRG}

In the present context, the functional renormalization group allows for the inclusion of fluctuations momentum-shell by momentum-shell. This is arranged by introducing a momentum cutoff $R_k$, which effectively suppresses fluctuations in spatial momenta $ p^2 \lesssim k^2$. The change of the effective action with lowering the cutoff scale $k$ is tracked using the Wetterich equation \cite{wetterich1993exact},
\begin{align}\label{eq:GenFlow}
	&\partial_\tau  \Gamma_k[\phi, \tilde \phi]= \frac{1}{2} \int_t \int_{p}\Tr \left[  G_{k}[\phi, \tilde \phi] \,\partial_\tau  R_{k}(p^2)\right] \,,
\end{align}
where the integral over momentum space is given by $\int_{p} = \int \frac{d^{d} p}{(2 \pi)^{d}}$ and the trace is evaluated over all internal field indices. Commonly, the equilibrium flows are evaluated in terms of frequency instead of time. However, with the present focus on non-equilibrium quantities we need to resolve the time-dependence, which is discussed further in \Cref{sec:TimeDep}.
We have also introduced the so-called RG-time $\tau =\ln(k/\Lambda)$ (not to be confused with real time $t$), where $\Lambda$ is some ultraviolet cut-off. 

Since both the fields $\phi$ and $\tilde{\phi}$ are $N$-components vectors evaluated on the static equilibrium configuration $\phi^*, \tilde \phi^*$, the Green functions as well as the regulator are $2N\times 2N$ matrices of block diagonal structure.
The propagator is simply given by the inverse of the RG-time dependent two-point function
\begin{align}\label{eq:prop}
	G_{k}[\phi, \tilde \phi] = \left[\Gamma_k^{(2)}[\phi, \tilde \phi] + R_k\right]^{-1} \,.
\end{align}
Furthermore, the response field formalism makes use of a block diagonal regulator matrix, which regulates the $\phi \tilde\phi$ contributions.
Each block of the regulator matrix is given by the matrix 
\begin{align}\label{eq:Regulator}
	R_{n,k}(p) =  p^2 K
	\begin{pmatrix}
		0 & r(y) \\[1ex]
		r(y) & 0
	\end{pmatrix}\,, \quad \mathrm{and} \quad y = \frac{{p}^2}{k^2}\,,
\end{align}
where we have omitted any time-dependence of the regulator function.
Presently, we use the Litim regulator shape function ~\cite{litim2000optimization, litim2001optimized}
\begin{align}\label{eq:LitimReg}
	r(y) = \left(1/y -1\right)\theta(1-y)\,,
\end{align}
We use a spatial cutoff function, which enables the evaluation of flows at constant spatial configuration/in momentum space, whilst considering the time-evolution of the field separately.

Precision calculations of critical exponents within the fRG need to include higher derivative terms in the effective action ansatz as well as explicit regulator optimization, see \cite{de2021precision} for a detailed discussion on precision equilibrium critical exponents calculations within fRG. However, the Litim regulator is known to perform well at low-order in derivative expansion justifying our current choice~\cite{defenu2018scaling}.

\section{Expansion around the homogeneous state}\label{sec:HomogeneousExp}

In principle, functional methods provide exact evolution equations, which allow for direct computation of time and spatially dependent expectation values/observables for any type of system or geometry. The crucial ingredient for such a computation is the approximation of the effective action, which needs to contain the relevant degrees of freedom.

However, a numerical evaluation of a system with a full temporal or spatial dependence is computationally expensive. Hence, we aim to set up a scheme which expands about a temporally and spatially homogeneous system. The underlying assumption is that at criticality the homogeneous solution already contains the relevant information, such that the non-equilibrium critical exponents can be easily extracted or obtained from a converging iteration procedure.

More concretely, we are interested setting up an expansion scheme which connects the short time aging behavior with the critical behavior appearing close to equilibrium, following~\cite{chiocchetta2016universal}. Since we are presently only interested in the time-dependence, we proceed to evaluate the fields on uniform spatial configurations. Accordingly, we may refer to time-dependent quantities as `inhomogeneous', in contrast to homogeneous, equilibrium ones.

For the aging behavior of model A, the system equilibrates at long time. Yet, at any finite $t$ the effective action will be minimized by a time-dependent state. Thus, it is possible to expand the out-of-equilibrium solution around the time translational invariant one at $t=\infty$. Consequently, all (possibly field dependent) couplings $\lambda \in \{Z, K, D, U(\phi), \dots\}$ within the effective action \labelcref{eq:ansatz1} can be decomposed into
\begin{align}
	\label{eq:gmto0}
	\lambda(\phi(t); t) = \lambda_{\rm hom}(\phi(t)) + \bar \lambda(\phi(t); t)\,, 
\end{align}
where the subscript $``\rm hom "$ denotes the homogeneous solution and $\bar \lambda$ is a correction that explicitly depends on time.

The second ingredient to the expansion is a time-dependent Gaussian propagator or Green's function, which has knowledge of the two-point couplings on the equations of motion:
\begin{align}\label{eq:hprop}
	G^0_{ii}(x, x'| \, m_i(t), Z(t), K(t), D(t)) \,,
\end{align}
where the $ii$ subscript distinguishes between the massive ($i=1$) and Goldstone ($i \neq 1$) cases, with $m_i = U^{(2)}_{ii}(\phi) \equiv \partial_{\phi_i}^2 U(\phi)$. Notice that non-diagonal propagator terms vanish within the ansatz in \labelcref{eq:ansatz1}. The Gaussian propagator is given analytically in \Cref{app:GaussProp} and is computed from the quadratic part of the effective action. In the $t_0\to -\infty$ limit we simply recover the equilibrium propagator, which is also reflected in \labelcref{eq:gmto0}.

In the following, we use the ansatz \labelcref{eq:gmto0} and systematically iterate the time dependent problem starting from the homogeneous (time-independent) solution, which is obtained at $t_{0}\to -\infty$. At leading order, the iterative procedure yields propagators whose dynamical structure is analogous to the one of the Gaussian propagators defined in \labelcref{eq:hprop} but where the quasi-particle spectrum $\omega_{q}$ is replaced with the one of the \textit{interacting} homogeneous solution. In the following, we will generically refer to this propagator as leading order propagator $G^{\rm LO}_{ij}(x,x')$.

\subsection{Diagrammatic structure}\label{sec:Diagrams}

We begin by extracting the non-Gaussian contribution $\Sigma$ to the propagator
\begin{subequations}
	\begin{align}\label{eq:split}
		G_{ij}(x,x')^{-1}=G^{\rm LO}_{ij}(x,x')^{-1}- \Sigma_{ij}(x,x')\,,
	\end{align}
    where $G^{\rm LO}$ displays the same time dependence as the Gaussian propagator in \labelcref{eq:hprop}.
Since our ansatz only features local interactions, see \labelcref{eq:ansatz1}, $\Sigma$ is given by
	\begin{align}\label{eq:V}
		\Sigma_{ij}(x,x')& =\Sigma_{ij}(x)\delta(x-x')=- \delta(x-x') \theta(t - t_0) \,\notag \\[1ex]
		&\hspace{-1.2cm} \times \begin{pmatrix}
			\tilde \phi_l U^{(3)}_{ijl} (\phi;t) &  U^{(2)}_{ij}(\phi; t)- U^{(2)}_{ij}(\phi^*;t)\\
			U^{(2)}_{ji}(\phi; t)- U^{(2)}_{ji}(\phi^*;t) & 0
		\end{pmatrix} \,,
	\end{align}
\end{subequations}
where $U$ is a fully field-dependent potential, which contains a time-dependence in the couplings, as well as the field $\phi(t)$. For the derivatives of the potential we are using the notation $U_{i_1,\dots,i_p}^{(p)} \equiv \partial_{\phi_{i_1}} \dots \partial_{\phi_{i_p}}U$. Furthermore, \labelcref{eq:V} only allows for a temporal inhomogeneity and is evaluated at spatially constant configurations. 

Once again, the reader shall notice that $G^{\rm LO}$ and $\Sigma$ are not the bare-propagator and self-energy of the system which are featured in the Dyson equation. Rather, they correspond to the aforementioned ingredients \labelcref{eq:gmto0} and \labelcref{eq:hprop}. 
The main advantage of the separation in \labelcref{eq:split} is splitting a known set of couplings \labelcref{eq:gmto0}, with no time dependence at leading order, and constructing their time evolution by convolving them with the explicit time-dependence of the Gaussian propagator. This constitutes the core of our expansion/iteration scheme.
The splitting \labelcref{eq:split} enables one to derive a Dyson-like identity, see Appendix C of \cite{chiocchetta2016universal},
\begin{align}\label{eq:Propexp}
	G_{ij}(x,x')=& G^{\rm LO}_{ij}(x,x')\nonumber\\
    &+ \int_{x''} G^{\rm LO}_{ik}(x,x'') \Sigma_{kl}(x'')G_{lj}(x'',x') \,,	
\end{align}
which is the starting point of the expansion scheme.

We can reinsert~\labelcref{eq:Propexp} into itself, creating a series in powers of the interaction term $\Sigma$. By dropping the last term of the series, we have created an expansion of the full propagator in terms of $G^{\rm LO}$ and $\Sigma$. We emphasize that $\Sigma_{kl}$ contains the full non-perturbative potential interactions of all orders, which distinguishes it from perturbative calculations.

For example, cutting the expansion at second order,
\begin{align}
	\label{eq:exp_gf}
	&G_{ij}(x,x')=G^{\rm LO}_{ii}(x,x')\delta_{ij}+\int_{y}G^{\rm LO}_{ii}(x,y)\Sigma_{ij}(y)G^{\rm LO}_{jj}(y,x')\notag \\
	&+ \sum_{l}\int_{y'}\int_{y}G^{\rm LO}_{ii}(x,y)\Sigma_{il}(y)G^{\rm LO}_{ll}(y,y')\Sigma_{lj}(y')G^{\rm LO}_{jj}(y',x')\,.
\end{align}
Inserting this expression for the propagator into the Wetterich equation is exact for the flow of the two-point function, but becomes inexact at the level of the four-point function. This is due to the property that the $n$-th order of the expansion generates all diagrams containing up to $n$ external vertices. Furthermore, there is no double counting, since the interaction term $\Sigma_{lj}$ is zero unless it is connected to at least one external line, i.e.~we have taken a derivative, by construction.

\subsection{Computation of time dependent couplings}\label{sec:TimeDep}

Our setup allows us to construct the time-dependence of any coupling in an iterative construction centered around the leading-order propagator:
\begin{enumerate}
	\item The iterative procedure begins setting $\bar \lambda^{n=0}(\phi; t) = 0$ in \labelcref{eq:gmto0} and computing the leading-order propagators based on the values of the homogeneous (time-independent) couplings.
	\item Then, the Dyson-like expansion in \labelcref{eq:Propexp} (at a given order $m$) is inserted into the Wetterich flow in \labelcref{eq:GenFlow} generating a set of equations to compute $\lambda(\phi; t)$. The resulting expressions for the couplings display explicit time dependence, which occurs due to the convolution of the homogeneous couplings with the (time-dependent) leading order propagator (see the last two terms in \labelcref{eq:exp_gf}). As a result, the first iteration of the procedure yields correction $\bar \lambda^{n=1}(\phi; t) \neq 0$.
	\item At any new iteration step $n$, time dependent corrections to the couplings $\bar \lambda^{n}(\phi; t) $ are extracted and  the scheme can be re-iterated until convergence is achieved.
\end{enumerate}
As pointed out in the previous section, the flow equations for the time dependent $\bar \lambda^n(\phi)$ are full functional equations, but are only exact up to the $m$-point functions, since diagrams are dropped in the expansion scheme afterwards. Therefore, it is essential that the Dyson-like equation~\eqref{eq:Propexp} is expanded to a high enough order $m$ so that one can obtain proper flow equations for all the non-trivial couplings considered in the initial ansatz.

At $n=0$, where the couplings are given by the homogeneous solution $\lambda_{\rm hom}$ the leading-order propagator in \labelcref{eq:hprop} is time-dependent. Thus, one can extract time-dependent quantities already at leading order, since most of the information concerning interactions is carried by the bulk homogeneous solution. From this perspective, the procedure introduced in \cite{chiocchetta2016universal} corresponds to performing the first item of the previous list.

In the present work, we again employ the leading order approximation, i.e.~step 1 of the aforementioned list. However, different from \cite{chiocchetta2016universal} we perform a fully functional solution of the bulk homogeneous couplings without further approximations with respect to the ansatz in \labelcref{eq:ansatz1}. 
Implementing the entire iteration procedure and the projection on the time-dependent coupling flows is deferred to future work. 

\subsection{The temporal boundary renormalization}

The renormalization of the temporal boundary is obtained from our ansatz \labelcref{eq:ansatz1}, by projecting on the two-point function
\begin{align}\label{eq:Z0Proj}
	Z_0 = \frac{ 1}{\mathcal{V}_d}\left. \frac{\delta^2 \Gamma_k[\phi, \tilde \phi]}{\delta \phi_{0,i} \delta \tilde \phi_{0,i}}\right|_{\phi(x) = \phi, \tilde \phi(x) = 0}\,.
\end{align}
Note, that even though $Z_0$ is a quantity related to the temporal boundary $t_0$, it is defined by the whole time-evolution thereafter. It is explicitly not a time-dependent coupling, but the normalization of the initial condition. This is reflected in the fact, that the $\phi_0$ derivatives are only functional derivatives in terms of spatial configurations.
Consequently, the time-direction has to be integrated fully in the computation of the flow. This is different in the computation of time-dependent couplings $\bar \lambda$ as suggested in \Cref{sec:TimeDep}. 
Lastly, $\mathcal{V}_d $ indicates the spatial volume.

To compute the flow on the boundary $Z_0$ we begin by inserting the expansion \labelcref{eq:exp_gf} into \labelcref{eq:GenFlow}.
The expression for the flow at spatially homogeneous field configurations $\phi(t, r) = \phi(t)$, $\tilde \phi(t, r) = \tilde \phi(t)$ and after performing the spatial/momentum integration reads
\begin{widetext}
	\begin{align}\label{eq:exempFlow}
		\frac{1}{\mathcal{V}_d} \partial_\tau \Gamma[\phi] =& - A_d k^{d+2}\left(1- \frac{\eta_K}{d+2}\right) \times  \int_{t'}  \tilde \phi_1(t') \Big\{  U^{(3)}_{111}(\phi; t')f_a(\omega_1; t') + (N-1) U^{(3)}_{122}(\phi; t')f_a(\omega_2; t') \notag\\[1ex]
		& 
		-\int_{t''} \sum_{i} U^{(3)}_{1ii}(\phi; t')\left(U^{(2)}_{ii}(\phi; t'')-U^{(2)}_{ii}(\phi^*; t'')\right)f_b(\omega_i,\omega_i; t', t'')
		\Big\} \notag\\[1ex]
		& - A_d k^{d+2}\left(1- \frac{\eta_K}{d+2}\right) \times  \int_{t'}  \tilde \phi_2(t') \Big\{  U^{(3)}_{211}(\phi; t')f_a(\omega_1; t') + (N+1) U^{(3)}_{222}(\phi; t')f_a(\omega_2; t') \notag\\[1ex]
		& 
		-\int_{t''}  2\left[ U^{(3)}_{221}(\phi; t')\left(U^{(2)}_{12}(\phi; t'')-U^{(2)}_{12}(\phi^*; t'')\right)f_b(\omega_1, \omega_2; t', t'')\right]
		\Big\}  \,.
	\end{align}
\end{widetext}
The terms $\propto  \tilde \phi_1 $ will contribute to the boundary using the projection onto the massive mode, whereas the terms $\propto  \tilde \phi_2$ give the Goldstone projection.
We have dropped all fields $\propto  \tilde \phi_n $ for $n>2$ as we do not project onto them and evaluate at $  \tilde \phi^*_{n\geq 2} = 0 $.
The anomalous dimension $\eta_K$ is given by the fixed-point value of
\begin{align}
	\eta_K  = - \frac{\partial_\tau K}{K} \,.
\end{align}
Lastly, the functions $f_a, f_b$ are given by integrals over the Gaussian propagators, for example
\begin{align}\notag \label{eq:explicitZ0}
	&f_{a}(\omega; t') = 2 \int_t G^{\mathrm LO}_{C}(t,t';\omega, \dots) G^{\mathrm LO}_{R}(t',t; \omega, \dots)  \\[1ex]
	&= \frac{D}{Z \omega^2} \left[1 - e^{- 2 (t'-t_0) \omega} \left(1 + 2 (t'-t_0) \omega\left[1 -\frac{Z \omega}{D \mu_0}\right]    \right)	\right]\,,
\end{align}
and $\omega = k^2 + m^2_i$ is the dispersion relation. The expression for $f_b$ can be found in \Cref{app:timeIntegrals}, where we also state the shape of the Gaussian propagators and give some more details.
The shape of $f_a$ and $f_b$ allows for the distinction of the equilibrium flow from the inhomogeneity introduced by the initial quench already at leading order of the proposed expansion scheme: As $t' \to \infty$ the $t'$ dependent part of the flow vanishes due to the exponential suppression factor and the remaining '1' is the full homogeneous flow (at the given expansion order of the flow equation). Similarly, the time-translation invariant limit, i.e.~the limit without temporal boundary, is regained by taking $t_0 \to - \infty$.

Computing the second line in \labelcref{eq:exempFlow} requires an additional approximation, since we have time-dependences in both $t'$ and $t''$. To compute the $t''$ integral, we perform an expansion of the field
\begin{align}
	\phi(t'') = \phi(t') + (t''- t') \dot \phi(t') \dots \,.
\end{align}
For the computation of the renormalization boundary $Z_0$ we may drop the second term, since we later evaluate at $\dot \phi(t_0)= 0$. However, this term is the leading contribution to the flow of the wave function of the time derivative $Z$ in \labelcref{eq:ansatz1}.

Finally, we remain with contributions of the general shape
\begin{align}
	A_i \propto  \int_{t'} \theta(t'-t_0) \tilde \phi(t') \,  g_i\left(U(\phi, t')\right) e^{- 2  \omega t'} \,,
\end{align}
for the inhomogeneous part of the flow. The last integral can be evaluated in terms of the boundary field by using the identity
\begin{align}
	\int_{t'}\theta(t'-t_0) g(t') e^{-c(t'-t_0)} = \sum_{n=0}^{\infty}  \frac{1}{c^{n+1}} \partial_{t}^n g(t)\vert_{t=t_0} \,,
\end{align}
where $c>0$ is a constant in time and corresponds to the homogeneous part of the dispersion relation in the present context.
This last step leaves us with an expression for the inhomogeneous flow of $\frac{1}{\mathcal{V}_d} \partial_\tau \Gamma[\phi] $ in terms of the initial fields $\phi_0$ and its time-derivatives.  

Now we use the equilibrium potential for the interaction vertex \labelcref{eq:V} as input. Since we are interested in the fixed-point equations, we insert the dimensionless fixed point values of the potential. Importantly, the pre-quench parameter $\mu_0$ does not receive any corrections from the flow and only runs with its dimension. Thus we can take the limit
\begin{align}
	\hat \mu_0 = \frac{\mu_0}{k^2} \to \infty \,,
\end{align}
at the fixed point.
Finally, the anomalous dimension at the boundary is given by
\begin{align}\label{eq:etaZ0}
	\eta_{Z_0} = -\frac{\partial_\tau Z_0}{Z_0}\,,
\end{align}
where the full expression in terms of the equilibrium fixed-point potential is indicated in \labelcref{eta0g}.
\begin{figure}
	\centering
	\includegraphics[width=\linewidth]{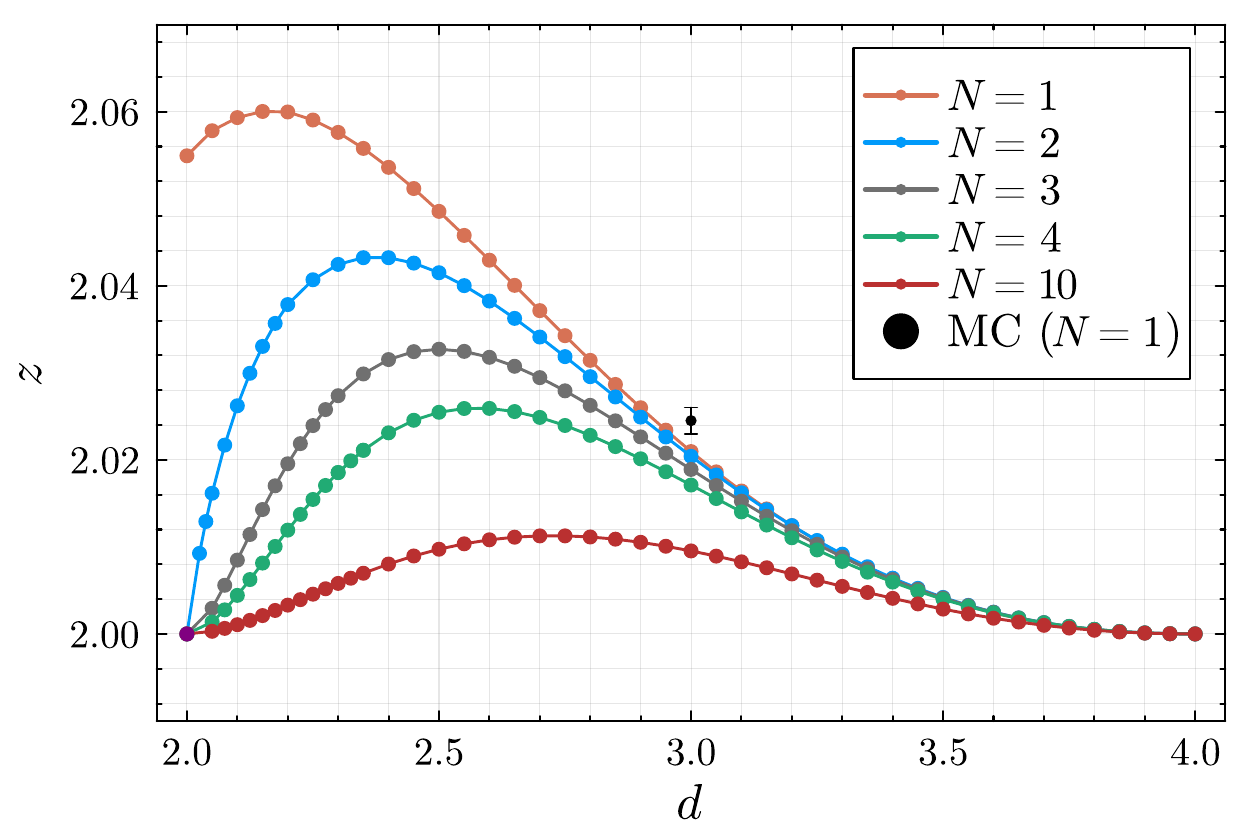}
	\caption{Dynamic exponent $z$ as a function of dimension and various $N$. The computation uses the equations stated in~\Cref{sec:PropsAndVerts}. The Monte Carlo (MC) point at $d=3$ is taken from~\cite{hasenbusch2020dynamic}. Furthermore, the MC point in $d=2$ is outside the range of the graphic ($z_{\text{MC},2} = 2.1667(5)$ from \cite{nightingale2000monte}) and is discussed further in \Cref{sec:results}. The purple point corresponds to $z=2$ at $d=2$, which is exact, as a consequence of the Mermin-Wagner theorem. 
 \hspace*{\fill}}
	\label{fig:equilibriumExp} 
\end{figure}

In summary, the leading order iteration uses the full equilibrium critical potential $U^*$ and the critical value of the anomalous dimension $\eta_K^*$ which are computed in \Cref{app:Numerics} in the conventional LPA' fashion for fixed-point equations of the $O(N)$ model~\cite{codello2013n,defenu2015truncation,codello2015critical}.
With this input data one can easily infer the critical value of $\eta_Z = - \partial_{t} \log (Z)$ and hence $z$~\cite{canet2007non,batini2023dissipation}, since the flow equation of $Z_k$ decouples from the system.
\begin{figure}
\includegraphics[width=\linewidth]{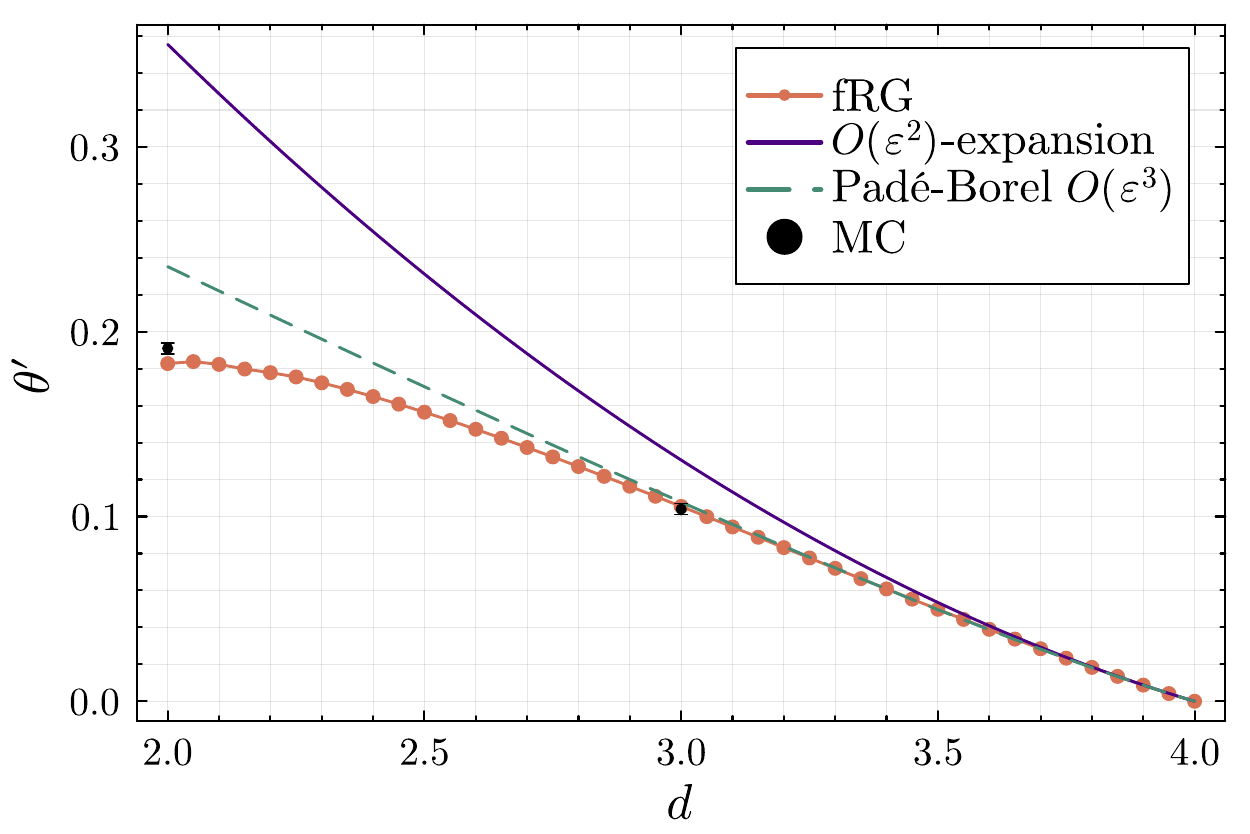}
\caption{Initial slip exponent $\theta'$ for $N=1$. We find that the homogeneous expansion within the fRG formalism yields estimates in agreement with the MC results for $d=2$ and $d=3$, taken from~\cite{grassberger1995damage}. As such, it carries an improvement over both the two-loop $\epsilon$-expansion (solid purple line) and the Padé-Borel resummation based on the three-loop $\epsilon$-expansion (dashed green line) found in~\cite{prudnikov2008renormalization}.
			\hspace*{\fill}}
		\label{fig:massiveO1}
	\end{figure}

\section{Results}\label{sec:results}
In the present work we numerically solve the equilibrium fixed-point equation in LPA' (see~\Cref{sec:PropsAndVerts}) using Chebyshev spectral methods, thus providing a full effective potential; more details on the numerical implementation are contained in~\Cref{app:Numerics}.
	
Based on this input, we compute the critical values of the anomalous dimension of the boundary renormalization $\eta_{Z_0}^*$, whose expression is given in \labelcref{eta0g}. 
Furthermore, we compute the equilibrium and close-to-equilibrium quantities $\eta_K^*$ and $\eta_Z^*$.
	
In fact, we begin by discussing the dynamic exponent $z = 2 - \eta_K^* + \eta_Z^*$. Our results for $z$ are plotted in~\Cref{fig:equilibriumExp} for several values of the number of components $N$ of the order parameter and for spatial dimensions $2<d<4$. 
We immediately observe that the well-known limiting cases are reproduced: $z = 2$ at the upper critical dimension $d=4$, independently of $N$, and again $z=2$ at the lower critical dimension $d=2$ for all theories with continuous symmetry ($N\geq 2$). The latter fact is in agreement with the Mermin-Wagner theorem~\cite{mermin1966absence}. On the other hand, in two dimensions $z > 2$ for the Ising case with $N=1$, as expected.

The fRG estimates of the critical exponents depend on whether the flow of the wave-function renormalization is defined by projecting on the Goldstone or massive two-point functions~\cite{defenu2020criticality}. This effect is crucial for obtaining accurate values of $\eta$ and $\nu$ in equilibrium calculations~\cite{codello2015critical}. However, we do not directly address this issue here. Instead, we employ the Goldstone definition for the computation of the anomalous dimension flow equations (see \labelcref{eta0m} and \labelcref{eq:etaIsing}) for all values of $N$, analytically continuing it to $N=1$.

In certain cases, such as the computation of the critical exponent $z$ in $d=2$, this approach may reduce our accuracy compared to other studies where the anomalous dimension definition is modified depending on the quantity under study~\cite{codello2015critical}. For completeness, the flow equations obtained with the massive definition are reported in \Cref{sec:O1equilibrium} and \labelcref{eta0g}.

Indeed, our estimate for $z$, see \Cref{fig:equilibriumExp} does not approach the MC result of Ref.~\cite{nightingale2000monte}, $z_{\text{MC},2} = 2.1667(5)$, in $d=2$. Nonetheless, we argue below that this does not affect significantly the results for the aging exponent.  At variance with the two-dimensional case, in $d=3$ we obtain $z = 2.021$ for the Ising model, which is in fair agreement with the recent MC  calculation $z_{\text{MC},3} = 2.0245(15)$ carried out in~\cite{hasenbusch2020dynamic}. 

Note that more technically involved fRG computations of $z$ using frequency-dependent regulator functions and the massive definition of the anomalous dimension further corroborate the MC simulations and produce results in the range $z_{\text{fRG}}=2.023-2.025$~\cite{mesterhazy2015quantum, duclut2017frequency}. Furthermore, another benchmark is provided by flattening the curves in~\Cref{fig:equilibriumExp} into $z(N\to \infty) = 2$ (uniformly in $d$) as $N$ increases. Indeed, in the large-$N$ limit the anomalous dimensions $\eta_K^*$ and $\eta_Z^*$ are both vanishing.

As anticipated, despite potential inaccuracies in the determination of the exponent $z$ -- i.e. in the anomalous dimensions -- the aging exponent, which is computed via
\begin{align}\label{eq:thetad}
\theta = \frac{\eta_{Z}^*-\eta_{Z_0}^*}{z} \,,
\end{align}
benefits from cancellations between both terms in the numerator of~\labelcref{eq:thetad}, which reduce the error in our estimate of $\theta$. It is therefore reasonable to expect quantitatively better results for the non-equilibrium exponents $\theta$ and $\theta'$ from our fRG approach.
\begin{figure}
\centering
\includegraphics[width=\linewidth]{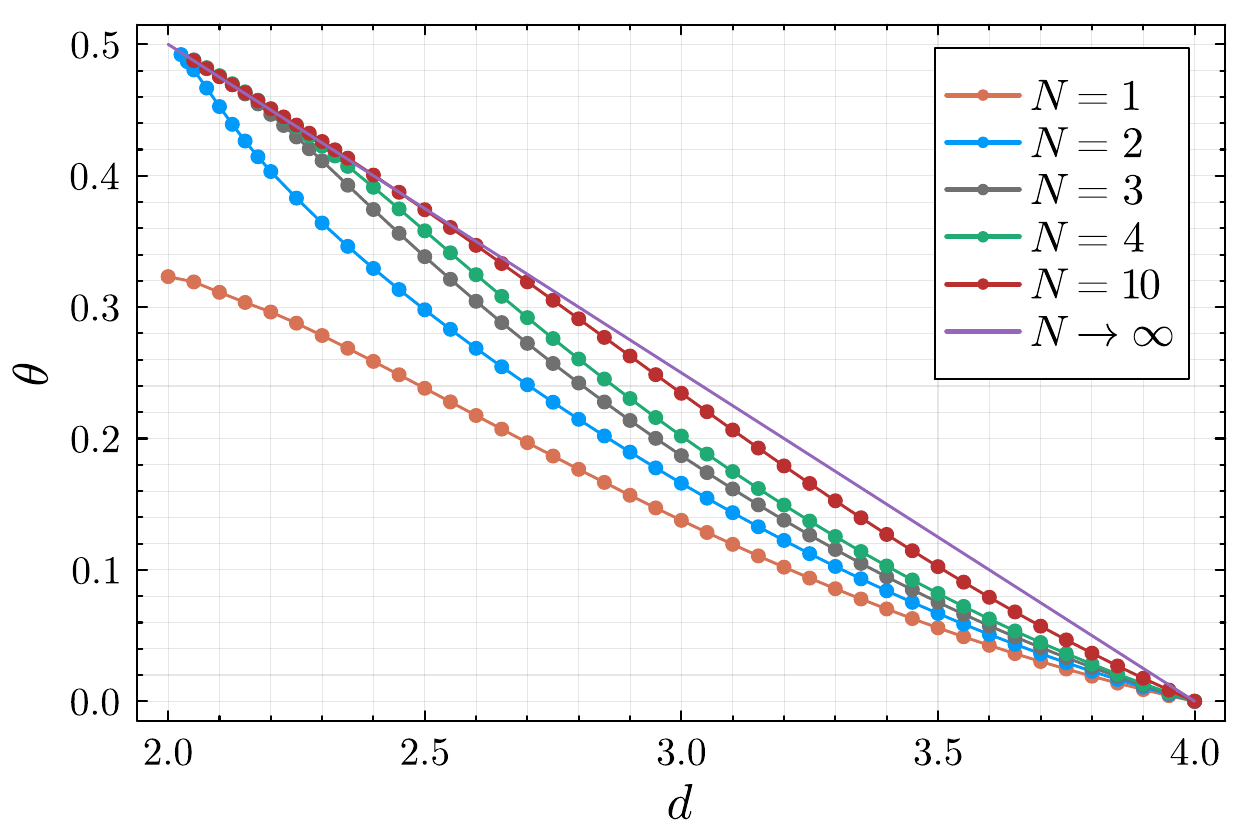}
\caption{Aging exponent $\theta$ as a function of dimension for various $N$ as computed from~\labelcref{eq:thetad}. The approach to the $N\to\infty$ exact result as the number of components increases is clearly observed. Moreover, in two dimensions $\theta'$ seems to converge to the value $1/2$, independent of $N$ ($>1$), as we discuss further in the text. 
  \hspace*{\fill}}
		\label{fig:generalON} 
	\end{figure}

In order to compare our calculations with other approaches in the literature focusing mostly on the initial slip exponent $\theta'$ of the Ising model -- as opposed to the aging exponent $\theta$, see \labelcref{eq:initial_slip_vs_aging} -- in~\Cref{fig:massiveO1} we show $\theta'$ as a function of dimension $d$ and $N=1$. For the first time we provide a truly non-perturbative prediction for the initial slip exponent in $d=2$. We also show two-loop and resummed three-loop $\epsilon$-expansion results~\cite{prudnikov2008renormalization} for comparison. Given the level of complexity of the homogenenous expansion, which is comparable with diagrammatics at two loops, the accuracy of our results is substantially better when compared to the MC estimates~\cite{grassberger1995damage}. In particular, the three-loop calculation is considerably more involved than our LPA' treatment, which shall be in turn more flexible and applicable also in presence of disorder, where the Borel-summability of the asymptotic series presents several difficulties~\cite{pelissetto2002critical}.
Therefore, our method shall provide a powerful alternative to perturbative approaches, especially in the out-of-equilibrium regime of systems with quenched disorder. In fact, the treatment of quenched random variables in the MSRJD formalism is similar to the one of thermal noise of the underlying Langevin dynamics. Quenched random variables can also be integrated out, at least in the approximation of Gaussian disorder, leading to an effective action with temporal non-locality.

As anticipated, we consider the MC results of~\cite{grassberger1995damage}, namely $\theta'_{\text{MC}} = 0.191(3)$ in $d=2$ and $\theta'_{\text{MC}} = 0.104(3)$ in $d=3$, which we have to compare with our fRG estimates 
\begin{equation}
    \theta' = 0.183 \quad \text{in $d=2$}, \qquad \theta' = 0.105 \quad \text{in $d=3$}
\end{equation}
We also note that subsequent MC simulations reported the following results: $\theta'_{\text{MC}} = 0.197(1)$ in $d=2$ from~\cite{okano1997universality} and $\theta'_{\text{MC}} = 0.108(2)$ in $d=3$ from~\cite{jaster1999short}. In any event, we can establish that the values of the initial slip exponent obtained via our fRG approach lie below the $5\%$ accuracy threshold.
	
Finally, the flexibility of our approach is demonstrated by the simplicity it generalizes to $N>1$, see~\Cref{fig:generalON}. As we noticed already for the dynamical exponent $z$, thanks to the non-perturbative nature of the fRG approach, the results for the aging exponent $\theta$ fulfill several important benchmarks: At $d=4$ we approach the Gaussian limit $\theta_{d=4}=0$. Moreover, for $N \to \infty$ the solution approaches the known large-$N$ limit $\theta=\epsilon/4$, where $\epsilon = 4-d$~\cite{janssen1989new}. Interestingly,  the fRG calculation yields $\theta_{d=2} = 1/2$ for all $N>1$. While $\theta_{d=2} = 1/2$ agrees with the $\theta=\epsilon/4$ for $N\to\infty$, our study suggest this to be true for all models with finite $N$ too, in analogy with the two-dimensional static exponents of the $O(N)$ models obtained in the $d=2+\epsilon$ expansion~\cite{brezin1976renormalization}. In particular, the anomalous dimension $\eta$ becomes independent of $N$ as we approach the limit $d=2$ where the kinetic content of the theory is trivial. However, the interpretation of the finite, $N$-independent, $\theta$ is not straightforward in the out-of-equilibrium framework. In order to test the validity of the $\theta_{d=2} = 1/2$ prediction, one would need to perform a numerical simulation of the quench from high-temperature to zero temperature, since $T_c \to 0$ linearly in $(d-2)$ as $d\to 2$ from above. This aspect warrants further investigation.
    
\section{Conclusion}\label{sec:conclusion}

Our work presents an iteration scheme to study quenches in many-body statistical systems. This scheme expands out-of-equilibrium propagators and vertices around the homogeneous solution attained as $t \to \infty $. At leading order, the propagators retain the dynamical structure of the Gaussian theory, but include the quasi-particle spectrum and masses of the interacting equilibrium solution. At higher orders, the time-dependent couplings can be computed iteratively by expanding the Wetterich equation in terms of classes of diagrams (see \Cref{sec:HomogeneousExp}). We have used a fully field-dependent fixed-point potential in LPA' as the only input, computed using pseudo-spectral methods.

This \textit{homogeneous expansion} offers a straightforward way to compute the aging exponent in a decoupled manner, allowing us to determine the renormalization boundary based purely on the analytically known time-dependence of Gaussian propagators and equilibrium vertices at leading order. A similar scheme was suggested in \cite{chiocchetta2016universal} within a smaller truncation. Our work builds on these ideas by employing the more extensive LPA' ansatz for the effective action. Additionally, we have integrated the setup into an iterative scheme to compute the full time-dependence of couplings to any order in the fields.
	
We have shown results at the leading order of the expansion for the $O(N)$ model as a function of dimension ranging from $d=2$ to $d=4$. Our calculations showed apparent convergence in the fully non-perturbative regime with $d<2.5$. This is a considerable improvement in comparison to the approach in \cite{chiocchetta2016universal} that was carried out only for the $O(1)$ case.
	
Results were compared to MC computations in two and three dimensions from the literature. Not only are the critical indices in $d=3$ relatively accurate, but even the behavior of the aging exponent in $d=2$ aligns with the expectations of numerical simulations. Importantly, the results meet the analytically known benchmarks, such as the large-$N$ limit and the weak-coupling limit close to $d=4$. The interesting behavior of the aging exponent towards $d=2$, converging to the $N$-independent value $\theta=1/2$ for all $N > 1$, also seems consistent with expectations. This finite $N$-independent result for continuous symmetries in $d=2$ is, to our knowledge, a feature of the out-of-equilibrium universality that had not been discussed before in the literature. It would be interesting to test its validity by performing numerical simulations in this scenario.

In summary, this work provides a basis for the computation of non-equilibrium critical exponents within a prototypical framework, specifically model A. Our results not only advance the understanding of critical dynamics but also open the door to more precise computational approaches: In fact the setup can be improved by using further iteration steps or even computing time dependent corrections using \textit{physics-informed} RG flows~\cite{ihssen2024physics}. In future work, we aim to refine our methodology by optimizing the choice of regulator~\cite{de2021precision}. 

Building upon our current framework, several compelling avenues for future research emerge. First, we emphasize that the same methodology paves the way for non-perturbative studies of critical phenomena in systems with a spatial boundary \cite{diehl1986field, diehl1997theory} -- rather than a temporal one -- by means of the fRG. We further envision an extension of the present method to the analysis of systems with both a spatial and a temporal boundary, via a proper enlargement of our truncation of the effective action, following \cite{marcuzzi2012collective}.

Returning to time-dependent criticality of systems eventually relaxing to equilibrium, it would be possible to explore other kinds of dynamics in the classification of Hohenberg and Halperin~\cite{hohenberg1977theory}. For instance, one could analyze model B where $z$ and $\theta$ are not independent of the equilibrium exponents and verify the relations $z=4-\eta$ and $\theta = 0$ \cite{calabrese2005ageing} with the fRG. Moreover, a non-perturbative analysis could be developed based on the homogeneous expansion to investigate the out-of-equilibrium dynamics of $O(N)$ models in the presence of quenched disorder, such as bond or site dilution~\cite{calabrese2005ageing}. Finally, recognizing the growing interest in systems with long-range interactions~\cite{defenu2023long}, we are actively working to extend our framework to encompass the dynamics of long-range $O(N)$ models~\cite{defenu2015fixed,defenu2017criticality,defenu2020criticality}.

On the other hand, recent studies on driven-dissipative systems have unveiled novel universality classes associated with non-equilibrium phase transitions. For instance, the onset of time-crystalline order in driven $O(N)$ models has been shown to exhibit unique critical behavior distinct from equilibrium scenarios \cite{daviet2024nonequilibrium}. Similarly, the investigation of critical exceptional points in non-equilibrium $O(N)$ models has revealed the emergence of several intriguing features \cite{zelle2024universal}. Extending our fRG approach to these contexts could provide a non-perturbative understanding of the critical dynamics at these genuine non-equilibrium fixed points.

Pursuing these directions will not only enhance our comprehension of non-equilibrium critical phenomena but also bridge the gap between theoretical predictions and experimental observations in complex many-body systems.

	\begin{acknowledgments}
		F.I. thanks Jan M. Pawlowski, J. Wessely for discussions. This research was funded by the Swiss National Science Foundation (SNSF) grant number 200021 207537, by the Deutsche Forschungsgemeinschaft (DFG, German Research Foundation) under Germany's Excellence Strategy EXC2181/1-390900948 (the Heidelberg STRUCTURES Excellence Cluster), the Swiss State Secretariat for Education, Research and Innovation (SERI). J.M. acknowledges the Pauli Center for hospitality. 
	\end{acknowledgments}
	
	
	\appendix
	\begingroup
	\allowdisplaybreaks
	
	\section{Equilibrium flow equations}\label{sec:PropsAndVerts}
		
	The equilibrium flows are derived by projecting onto the corresponding tensor structures in the effective action and by evaluating on homogeneous field configurations.
	In particular we use the $O(N)$ invariant on the equations of motion
	\begin{align}
		\rho = \frac{\left(\phi^*\right)^2}{2} \,.
	\end{align}
	The flow of the potential and the equilibrium anomalous dimension $\eta_K$ in model A are equivalent to those of a static $O(N)$ model, which was already pointed out in \cite{duclut2017frequency}.
	The equilibrium $O(N)$-universality classes have been studied in arbitrary dimension in \cite{codello2013n,defenu2015truncation,codello2015critical} and the derivation of the equations is given in detail in \cite{berges2000nonperturbative}.
	
	In the following we use the notation $u^{(1)}(\rho) = \partial_{\rho}U/k^2$ for the dimensionless potential. Note, that the derivatives in the super-script are taken with respect to the invariant $\rho$. The anomalous dimensions are given by
	\begin{align}
		\eta_A = -\frac{ \partial_\tau A}{A} \,,
	\end{align}
	where $A= K,Z,D$.
	Finally, the dimensions of the potential and fields are given by
	\begin{align}\label{eq:dimensions}
		\hat \rho &= \frac{\rho}{k^{d_\rho}}\,, \quad d_\rho = d-2 \,, \notag \\[1ex]
		u^{(1)} &= \frac{\partial_\rho U}{k^{d_u}}\,, \quad d_u = 2\,.
	\end{align}
		\begin{widetext}

		\subsection{Flows in the $O(N)$ model}\label{sec:ONequilibrium}
		
		We list the flows by considering projections onto the massless mode, as is commonly done for $N>1$. All results for $N>1$ are generated from these equations. The flow of the dimensionless derivative of the effective potential is given by
		\begin{align}\label{eq:dtU}
			k \partial_k  u^{(1)} + (2 - \eta_K) u^{(1)} -(d-2 + \eta_K) \hat \rho u^{(2)}  =&  A_d \frac{D}{Z} \left(1 - \frac{\eta_K}{d+2}\right)
			\partial_{\hat \rho}\left\{ \frac{1}{1 + u^{(1)}+ 2 \hat \rho u^{(2)}} +  \frac{N-1}{1 +  u^{(1)}} \right\}\,,
		\end{align}
		where $A_d = \frac{2 \pi^{d/2}}{\Gamma(d/2) (2 \pi)^d d}$ originates from the angular momentum integration. The remaining terms are obtained from the two-point function of the Goldstone modes
		\begin{align}\notag
			\eta_K 	=& 4 A_d \frac{D}{Z} \frac{ \hat \rho \left( u^{(2)}\right)^2}{(1 + u^{(1)}+ 2 \hat \rho u^{(2)})^2 (1 + u^{(1)})^2}\,,\\[1.5ex]
			\eta_Z   =& A_d \frac{D}{Z} 
			\hat \rho \left(u^{(2)} \right)^2 \left(1-\frac{\eta_K}{d+2}\right)
			\frac{6 +(u^{(1)})^2 + 4 u^{(1)}(u^{(1)}+ 2 \hat \rho u^{(2)})+( u^{(1)}+ 2 \hat \rho u^{(2)})^2+12  \left(u^{(1)}+  \hat \rho u^{(2)}\right)}{(1 + u^{(1)}+ 2 \hat \rho u^{(2)})^2(1 + u^{(1)})^2(1 + u^{(1)}+  \hat \rho u^{(2)})^2}\,,
		\label{eq:eta}
		\end{align}
	and lastly $\eta_D = \eta_Z$ as required by detailed balance.
	In our calculation we use $D=Z$ at the fixed-point, such that their ratio drops from the flows.
	
	\subsection{Flows in $O(1)$}\label{sec:O1equilibrium}
	
	We also indicate the projections onto the massive mode for the $N=1$ computation.	
	The potential is given by
		\begin{align}
			k \partial_k  u^{(1)} + (2 - \eta_K) u^{(1)} -(d-2 + \eta_K) \hat \rho u^{(2)}  =&  A_d \frac{D}{Z} \left(1 - \frac{\eta_K}{d+2}\right)
			\partial_{\hat \rho}\left\{ \frac{1}{1 + u^{(1)}+ 2 \hat \rho u^{(2)}} \right\}\,.
		\end{align}
		In distinction to $N>1$ we obtain the wave function by projecting onto the massive mode.
		Then the anomalous dimensions read
		\begin{align}
			\eta_K 	=& 2 A_d \frac{D}{Z} \frac{ \hat \rho \left(3  u^{(2)}+ 2 \hat \rho u^{(3)}\right)^2}{(1 + u^{(1)}+ 2 \hat \rho u^{(2)})^4}\,, \hspace{2cm}
			\eta_Z  = 3 A_d \frac{D}{Z} \left(1-\frac{\eta_K}{d+2}\right)\frac{ \hat \rho \left(3  u^{(2)}+ 2 \hat \rho u^{(3)}\right)^2}{(1 + u^{(1)}+ 2 \hat \rho u^{(2)})^4}\,.
			\label{eq:etaIsing}
		\end{align}
	\end{widetext}

\section{Numerical evaluation of fixed-point equations}\label{app:Numerics}

This Appendix describes the numerical procedure of evaluating the full fixed-point solution in LPA'. 
We begin by discussing the standard ``spikeplot'' approach in \Cref{sec:Spikeplot}. In \Cref{sec:PseudoSpectral} we detail the pseudo-spectral method employed in the present work, emphasizing the advantages of the latter.

\subsection{Spikeplot approach}\label{sec:Spikeplot}

Given the ansatz \labelcref{eq:ansatz1}, the determination of the fixed-point behavior of the theory entails the solution of the ordinary differential equation (ODE) obtained by setting $\partial_\tau u^{(1)} = 0$ in \labelcref{eq:dtU}, where $\eta_K$ is given by \labelcref{eq:eta}. Typically, in order to numerically integrate the ODE, one uses a shooting method known as ``spikeplot" \cite{morris1994truncations,codello2012scaling,hellwig2015scaling,defenu2018scaling}. 

Due to the peculiar structure of the second order fixed-point ODE, the initial condition $u^{(2)}(0)$ depends on $\sigma \coloneqq u^{(1)}(0)$, as one can check by setting $\hat\rho = 0$ in the equation. Hence, we only need to determine the value of $\sigma$ to obtain a well-defined initial value problem. The criterion for correctly choosing the initial condition comes from the observation that a generic $\sigma$ causes the solution $u^{(1)}(\hat \rho)$ to be singular at a finite value of the field, which we call $\hat\rho_\infty(\sigma)$. Since singular solution are not physically acceptable free energies\,\cite{gallavotti2013statistical}, the condition $\hat\rho_\infty(\sigma) \to \infty$ (numerically, it means that we look for a \textit{spike} in the plot $\hat\rho_\infty$ vs $\sigma$) corresponds to requiring a physical fixed point solution: This produces a finite set $\{\sigma_{*,j}\}$. Each $\sigma_{*,j}$ corresponds to a different equilibrium fixed point (Gaussian, Wilson-Fisher, etc.) whose effective potential is obtained upon solving the initial value problem with initial condition $\sigma_{*,j}$.

A disadvantage of the spikeplot method is that, since it is initially unknown, the anomalous dimension $\eta_K$ has to be determined self-consistently. Without entering into the details, this involves the choice of a suitable initial guess for $\eta_K$ and a mixing procedure to update its value until convergence is reached, which is not guaranteed.

\subsection{Pseudospectral method}\label{sec:PseudoSpectral}

In most cases, however, the pseudospectral collocation method, which was introduced for RG fixed-point equations in \cite{borchardt2015global}, offers a more convenient alternative. We will now briefly describe how it works; for a more detailed introduction see e.g. \cite{boyd2001chebyshev}.

Consider a generic ODE
\begin{equation}\label{eq:generic_ODE}
    \mathcal{D}[y(x)] = 0,
\end{equation}
where $\mathcal{D}$ is a (non-linear) differential operator and $x \in [a,b]$ with finite $a$ and $b$ (later we discuss how to extend the discussion to our case with $\hat\rho \in [0,\infty)$). The core of the collocation method is to convert the ODE into a system of algebraic equations obtained by enforcing \labelcref{eq:generic_ODE} at specific \textit{collocation points}. First, we approximate $y$ using a basis of orthogonal polynomials. Our choice is to consider Chebyshev polynomials $T_n(s)$, where $s \in [-1,1]$, because of their advantageous convergence properties \cite{borchardt2015global,boyd2001chebyshev}. Then $y$ is replaced by
\begin{equation}
    y_{\text{approx}}(x) = \sum_{n=0}^{N_\text{coll}} y_n T_n\left(\frac{2x-a-b}{b-a}\right)\,,
\end{equation}
for some finite $N_\text{coll}$. The determination of the $N_\text{coll}+1$ coefficients $\{y_n\}$ would then yield the approximate solution to the ODE. This is obtained by evaluating \labelcref{eq:generic_ODE} with $y \rightsquigarrow y_{\text{approx}}$ at $N_\text{coll} + 1$ collocation points, which can be chosen as the extrema of the Chebyshev polynomial of order $N_\text{coll}$, i.e.
\begin{equation}
    s_j = - \cos\left(\frac{j \pi}{N_\text{coll}}\right) \in [-1,1], \qquad j = 0, \dots, N_\text{coll},
\end{equation}
to be mapped to the interval $[a,b]$. As anticipated, this procedure yields a system of $N_\text{coll}+1$ \textit{algebraic} equations in $N_\text{coll}+1$ unknowns, which can be numerically solved. It is however important to choose an initial guess for the Chebyshev coefficients $\{y_n\}$ that allows us to converge to the desired solution (in our context, among the several ones corresponding to different fixed points of the RG).

Possible boundary conditions can be explicitly imposed at the endpoints and added to the system of equations. Then, the set of collocation points must be appropriately modified to exclude $x=a,b$. 

As already discussed, our ODE of interest, with $y \equiv u^{(1)}$ and $x \equiv \hat\rho$, is such that only one boundary condition has to be specified. It turns out that this condition is automatically implemented by the collocation method.

So far, $a = 0$, but $b$ was not specified. Numerically, one could take $b$ to be a large value $\hat\rho = \hat\rho_\text{L}$. Alternatively, we can keep $b$ relatively small (but larger than the minimum of the potential) and solve the ODE on the semi-infinite interval $[0,b] \cup [b,\infty)$ by using the rational Chebyshev polynomials (see \cite{borchardt2015global}). This has the additional advantage that we can explicitly impose the large-field behavior of the solution, that is $u^{(1)}(\hat\rho \gg 1) \sim \hat\rho^{(2-\eta_K)/(d-2+\eta_K)}$.  

At variance with the spikeplot approach, the calculation of $\eta_K$ needs no iterative procedure: The algebraic equation for $\eta_K$ in \labelcref{eq:eta} is just added to the system of equation obtained by the collocation method. In other words, the determination of the effective potential and anomalous dimension is simultaneous.

In this work, the pseudo-spectral approach was preferred to obtain the critical exponents for all values of $d$ and $N$, due to its robustness and speed over the shooting method. Nonetheless, in the models with continuous symmetries $N>1$ at very low dimensions $d \gtrsim 2$ the minimum of the potential $\hat\rho_0$ becomes very large and diverges at the lower critical dimension $d=2$, where spontaneous symmetry breaking is absent \cite{mermin1966absence,defenu2015truncation}. In these cases the spikeplot approach is much easier to implement, since it does not rely on the preliminary choice of a finite interval $[a,b]$. Therefore, we have used that to initialize the Chebyshev coefficients and subsequently solve the system of non-linear algebraic equations obtained by collocation.

While in most cases it is more convenient to employ the pseudo-spectral collocation procedure, the combination of the two different approaches allows us to achieve great flexibility and precision in the computation of critical exponents in the whole space of $d$ and $N$.
	
	\section{Time evolution}\label{app:GaussProp}

In this Appendix we indicate the analytic shape of the Gaussian propagators, as well as the analytic expression for the anomalous dimension of the renormalization boundary.

\subsection{Gaussian propagators}
\label{g_prop}
In the absence of interactions ($g=0$ in \labelcref{eq:Hamiltonian}) the time evolution \labelcref{eq:LangevinEQ} is linear and can be solved analytically. The steps are outlined in \cite{chiocchetta2016universal}, and we simply state the result for our truncation of the effective action \labelcref{eq:ansatz1} and \labelcref{eq:InitialC}
	\begin{align}\label{eq:GaussianProps}
		G_{R}^0(q, t_1, t_2) =& \frac{\theta(t_1-t_2)}{Z} \exp(-  \omega_{q}(t_1-t_2)/Z) \,, \notag \\[1ex]
		G_{C}^0(q, t_1, t_2) =& \frac{D}{Z \omega_{q}}\Bigg[\exp(- \omega_{q}|t_1-t_2|/Z)  \notag \\[1ex]
		&\hspace{-2cm}+ \frac{ Z_0^2 }{Z^2} \left(\frac{\omega_{q} Z}{D \mu_0}+1-2 \frac{Z}{Z_0}\right)\exp(- \omega_{q}(t_1+t_2-2 t_0)/Z) \Bigg]\,,
	\end{align}
	where $G^0_{R}$ is the free response function and $G^0_{C}$ the free correlation function.
	At the zeroth order of the iteration procedure \Cref{sec:TimeDep} we set $D=Z=Z_0$, as they have not received any non-homogeneous corrections yet.
	
	The equilibrium dispersion relation is given by
	\begin{align}
		\omega_{q, n} =  m^2_{ \phi^n} + K q^2 \left[1 + r(q^2/k^2)\right] \,.
	\end{align}
	Finally, the free propagator matrix is then summarized as
	\begin{align}\label{eq:G0form}
		G^0(t_1,t_2) = 
		\begin{pmatrix}
			G^0_{C}(q, t_1, t_2) & G^0_{R}(q, t_1, t_2)  \\[1ex]
			G^0_{R}(q, t_2, t_1)  & 0 
		\end{pmatrix}\,.
	\end{align}
	The response function vanishes for $t_1<t_2$ due to the Heaviside theta function. This is the direct implementation of causality in the present setup. To manifest this further we also enforce $\theta(0)=0$ in the present setup, which ensures that the response function can only act on $t_1>t_2$.

	\subsection{Time integration}\label{app:timeIntegrals}
	
	At the leading order of the iteration scheme outlined in \Cref{sec:TimeDep}, all time-integrals can be performed analytically. The main text discusses the shape of the threshold function for tadpole diagrams \labelcref{eq:thetad}. Each additional external vertex requires one time integration more. For example the two external vertex type diagrams are described by the threshold function
	\begin{align}\notag
		f_{b}(\omega_1,\omega_2; t', t'') &= \int_t \left\{
		G^{\rm LO}_{C}(\omega_1; t',t'')\times\right. \notag \\[1ex]
		&\hspace{-2.5cm}\left[ G^{\rm LO}_{R}(\omega_2; t,t'') G^{\rm LO}_{R}(\omega_2; t',t)
		+ G^{\rm LO}_{R}(\omega_2; t,t') G^{\rm LO}_{R}(\omega_2; t'',t)\right]\notag \\[1ex]
		&\hspace{-2cm}+G^{\rm LO}_{C}(\omega_1; t,t')\times\left[ G^{\rm LO}_{R}(\omega_2; t',t'') G^{\rm LO}_{R}(\omega_1; t'',t) \right. \notag \\[1ex]
		&\left.+ G^{\rm LO}_{R}(\omega_1; t'',t) G^{\rm LO}_{R}(\omega_2; t'',t')\right] \notag \\[1ex]
		&\hspace{-2cm}+G^{\rm LO}_{C}(\omega_1; t'',t)\times \left[ G^{\rm LO}_{R}(\omega_1; t',t) G^{\rm LO}_{R}(\omega_2; t'',t') \right.\notag \\[1ex]
		&\left.\left.+ G^{\rm LO}_{R}(\omega_1; t',t) G^{\rm LO}_{R}(\omega_2; t',t'')\right]
		\right\} \,.
	\end{align}
	The integration can be performed analytically with e.g.~Mathematica.

	Together with the threshold function $f_a$ from the main text, the projection \labelcref{eq:Z0Proj} can be used to derive the explicit expression for the anomalous dimensions in terms of the full fixed-point potential used in \Cref{sec:PropsAndVerts}. 
	
	For the massive and Goldstone modes they read respectively
	\begin{widetext}
		\begin{align}
			\eta_{Z_0, 1}= &-A_d \frac{\hat D_k}{\hat Z_0}\left(1- \frac{\eta_K}{d+2}\right) \times    \left\{ \frac{ 3 u^{(2)} + 12 \hat \rho u^{(3)}+ 4 \hat \rho^2  u^{(4)}}{\hat \omega_1 ^3} + (N-1) \frac{u^{(2)}+2 \hat \rho  u^{(3)}}{\hat \omega_2^3} \right. \notag\\[1ex]
			& \hspace{4cm} \left. 
			-3 \hat{\rho} \left[ \frac{ \left(3 u^{(2)} + 2 \hat \rho u^{(3)}\right)^2}{\hat \omega_1 ^4}+ (N-1) \frac{\left(u^{(2)}\right)^2}{\hat \omega_2^4}\right]
			\right\} \,, \label{eta0m} \\[1ex]
			\eta_{Z_0, 2}= &-A_d \frac{\hat D_k}{\hat Z_0}\left(1- \frac{\eta_K}{d+2}\right) \times  \left\{  \frac{ u^{(2)} + 2 \hat \rho u^{(3)}}{\hat \omega_1^3} + (N+1) \frac{u^{(2)}}{\hat \omega_2^3}
			- 4 \hat \rho  
			(u^{(2)})^2  \frac{\hat \omega_2^4 + 2 \hat \omega_2^3 \hat \omega_1  +  2 \hat \omega_2 \hat  \omega_1^3 +  \hat \omega_1^4 }{\hat \omega_1^3 \hat  \omega_2^3 ( \hat \omega_2 + \hat \omega_1 )^2}
			\right\} \,.\label{eta0g}
		\end{align}
	\end{widetext}
	%


	\bibliography{ref_lib}
	
\end{document}